\documentclass[]{aa}
\onecolumn
\input epsfig.sty

\def\rw{{_{(0)}}}

\def\ini{{(ini)}}
\def\AA{{\it 3}}

\def\density{{\rho}}

\def\x{{\Delta}}
\def\y{{\Theta}}
\def\rr{{\delta\rho\over \rho\rw}}
\def\ss{{\delta\vartheta\over \vartheta\rw}}
\def\ix{{\mu}}
\def\Vp{{\em V}}
\def\dVp{{\em \delta V}}
\def\Vp{{V}}
\def\dVp{{\delta V}}

\def\Lorentz{{\left({1\over\density} 
	(H^i H_j -{1\over2} H^2\delta^{i}_{j})^{;j}
	      -{1\over\density}p^{;i}\right)_{;i}}}

\def\RayE{{Raychaudhuri-Ehlers }}

\def\ksi{{\left(\sqrt{\frac{1+\red_\ini}{1+\red}}\right)}}
\def\red{{\rm z}}
\def\MM{{M}}
\def\pan{{\em pancake }}
\def\wnr{{\em weakly nonlinear regime}}

\begin{document}

\title{Magnetized cosmic walls}

\author{Gra\.zyna Siemieniec-Ozi\c{e}b{\l}o
\and Andrzej Woszczyna}

\institute{Astronomical Observatory, Jagellonian University\\
Faculty of Mathematics, Physics and Computer Science\\
ul. Orla 171, 30--244 Krak\'ow, Poland\\
e-mail: grazyna\@@oa.uj.edu.pl}
\vspace{30mm}
\date{Received 12 March 2002 / Accepted 19 September 2003 }
\titlerunning{Magnetized cosmic walls}
\authorrunning{G. Siemieniec-Ozi\c{e}b{\l}o and A. Woszczyna}

\abstract{Nonlinear growth of one-dimensional density structures with a frozen-in magnetic field is
investigated in Newtonian cosmology. A mechanism of magnetic field amplification is
discussed. We discuss the relation between the initial conditions 
for the velocity field and the basic time-scales determining the growth of the magnetized structure.}
 
\maketitle


\section{Introduction}
During the last decade the evidence large-scale cosmological magnetic fields has systematically grown. Fields of  several microgauss have been measured beyond the galaxy
clusters (Kim et al. 1991); the Rotation Measure (Kronberg 1994) confirms the existence of coherent
magnetic field on Mpc scales or larger. The recent discovery of large-scale diffuse radio emission
testifies to the presence of magnetic fields of $\sim 0.1~\mu $G, along the 6 Mpc filament (Bagchi et al. 2002), 
with evidence for their coherent nature.

Understanding of the magnetic field behaviour at different phases of the matter-dominated era is crucial
to explain the microgauss fields observed in high redshift objects  $z \geq 2$; starting from the
damped Ly $\alpha$ systems, through the distant radiogalaxies and the galaxy clusters up to the scales
typical of galaxy superclusters. While the magnetic field of galaxies may result from dynamo effects, the
fields at larger scales cannot be explained by the same mechanism.  Here there is either no 
rotation, necessary for dynamo action, or the structures  are dynamically too
young to leave dynamo action enough time to operate. Magnetic fields at Mpc scales are likely to be
primordial. Some pre-dynamo mechanisms of
the primeval magnetic field amplification must be at work at least in the linear and {\wnr}.

In this paper we discuss  the mutual relationships between density growth and the magnetic field
evolution in early nonlinear stages, when the wall or filamentary structure is formed. We investigate
whether the growing planar density structure may drag and amplify the magnetic field.  We employ the exact
solutions for 2-D (\pan) inhomogeneity evolution in the Newtonian description\footnote{The problem
is opposite to that formulated by Wasserman (1978) and Kim et al.(1996), where the magnetic fields are
expected to actively support the structure formation processes. } and emphasize
the role of initial conditions, in particular, the large-scale primordial flows. The magnitude of
primordial velocity fields at recombination determines the time and the growth rate of density
fluctuation. As a consequence, it defines the duration time of the pre-dynamo and dynamo amplification phase.
To avoid problems with a mathematical 
definition of the {\wnr} we work with fully nonlinear equations and their solutions. Although finally we
refer to the regime where the density contrast $\x$ is between $1$ and $100$ (which is relevant
for the cosmological structures we discuss), dynamical equations are true for $\x>100$.

Magnetohydrodynamic equations in the covariant notation are given in Section~\ref{magneto}. Simplifying  physically 
relevant assumptions and the resulting nonlinear perturbation equations are discussed in
Section~\ref{planar}. Nonlinear solutions for the density contrast and the magnetic field enhancement
are given in Section~\ref{evolution}. Section~\ref{numerical}  contains numerical estimations and
graphical presentation of the magnetized \pan formation.

\section{Magnetized self-gravitating fluid}
\label{magneto}

Magnetohydrodynamic equations for self-gravitating fluid with infinite conductivity 
(Chandrasekhar $\&$ Fermi 1953, Wasserman 1978, Papadopulos $\&$ Esposito 1982), expressed
in a manifestly covariant way, form a dynamical system for the density $\density$, the
expansion rate $\vartheta$, and the magnetic field $H^i$
\begin{eqnarray}
\dot{\density} &=& - \density \vartheta\label{cg1}\\
\dot{\vartheta} &=& -{1 \over 3}\vartheta^2 -{4\pi}\kappa\density - 2\sigma^2+ 2\omega^2 +\Lorentz \label{ray1}\\
\dot{H^i}&=&\sigma^i_{j}H^j+\omega^i_{j}H^j-{2\over3}\vartheta H^i \label{ind1}
\end{eqnarray}
\noindent
Hydrodynamic scalars $\sigma$ and $\omega$, which measure the shear and the rotation, 
respectively, 
\begin{eqnarray}
\sigma&=&{1\over 2} (\sigma_{ik} \sigma^{ik})^{1/2},~~~~~~\sigma_{ik} ={1\over 2} (v_{i;k}+v_{k;i}-{2\over 3} \delta_{ik}v^ j_{;j}),\\
\omega&=&{1\over 2} (\omega_{ik} \omega^{ik})^{1/2},~~~~~~\omega_{ik} ={1\over 2} (v_{i;k}-v_{k;i}).
\end{eqnarray}
$v^i$ stands for the fluid velocity, its divergence builds the expansion rate
$\vartheta=v^i_{;i}$. Semicolon means the covariant derivative with respect to the space
coordinates, dot stands for the time convective derivative, e.g.
$\dot{\density} ={\partial \density/\partial t+v^i \density_{;i}}$, while $\delta^{ij}$
is the Kronecker delta; the Einstein summation convention is employed. Equations
(\ref{cg1}-\ref{ind1}) are coordinate-independent. (In the Cartesian coordinates, 
covariant derivatives reduce to the partial derivatives, and tensor indices are raised
and lowered by Kronecker delta). Equation (\ref{cg1}) is the continuity equation.
Equation (\ref{ray1}) is a Newtonian analogue of the \RayE equation (Ellis 1971) --- a scalar
form of the equation of motion for continuously gravitating media with the Poisson equation
included. System (\ref{cg1}-\ref{ind1}) has the same form in general relativity and
Newtonian theory (Ellis 1971, compare also Tsagas $\&$ Barrow 1997), and therefore is of particular
interest in the cosmological context.

The approximation of infinite conductivity results in a vanishing electric field. The
magnetic contribution to the equation of motion is quadratic in $H^i$ (the last term in
eq. (\ref{ray1})), and therefore, can be neglected in the weak magnetic field limit. In
this limit the system (\ref{cg1}-\ref{ind1}) splits into the autonomous system
(\ref{cg1}-\ref{ray1}) and the induction equation (\ref{ind1}). Then, the fluid
dynamics is entirely determined by gravitational forces, while the magnetic field is
dragged along by fluid flow.

\section{Planar symmetry}
\label{planar}

Below we consider a model based on the following assumptions:

1. The unperturbed ``background space" is an isotropic and homogeneous (Newtonian)
universe.

2. The initial perturbation state is given at random, i.e. perturbation in the density and velocity fields 
$\x_\ini=\left[\rr\right]_\ini$ and $\y_\ini=\left[\ss\right]_\ini $
are independent  quantities. The perturbation is initially small, i.e.
$\x_\ini\ll 1$ and $\y_\ini \ll 1$ 

3. After the recombination epoch, the infinite conductivity approximation is adequate.
The electric field vanishes ($E^i\simeq0$), while the magnetic field $H^i$ is frozen  
(Chandrasekhar $\&$ Fermi 1953). The matter pressure is negligible ($p= 0$)

4. During the considered period the primordial magnetic fields and their gradients are small 
compared with the density and the density gradients, respectively ($H_i H^i \ll \density$,
$H_{i;j} H^i \ll \density_{;j}$).  In the noncovariant approach to MHD this implies a small value of 
the ratio of Lorentz force $F_L$ to the gravitational force $F_G$.\footnote{
We have $F_L/F_G \propto \frac{H^2}{l (\rho + \delta \rho) G \rho l} \propto \frac{H^2}{\delta \rho} (\frac{t_{coll}}{l})^2$, where l is the typical length scale and $t_{coll}$ --- the characteristic time scale of collapse. Expressing then the collapse time by the quantity used in the above notation, $t_{coll} \propto \frac{1}{\delta \theta}$, one obtains for the ratio: $\frac{H^2}{\delta \rho} (\frac{1}{l \delta \theta})^2 \ll 1$ 
for all relevant Mpc scales of cosmological structures.}

5. The perturbation is rotationless  and has a planar symmetry, i.e. the velocity potential $\Vp (t,x^i)$ can
be expressed as $\Vp(t,x^i)=\Vp_{0}(t,x^i)+\dVp(t,x^3)$, where
$\Vp_{0}(t,x^i)={1\over 6}\vartheta\rw x_k x^k$ stands for unperturbed Hubble flow, and
the perturbation $\dVp(t,x^3)$ is independent of two of Cartesian coordinates.
Consequently, all the hydrodynamic scalars depend solely on time and on the only one
space variable $x^3\equiv z$ , the one parallel
to the fluid contraction (orthogonal to the \pan plane). 


Under these assumptions the system (\ref{cg1})-(\ref{ind1}) can be divided into the
background dynamics
\begin{eqnarray}
\dot{\density}\rw& =& \density\rw \vartheta\rw\label{cgRW}\\
\dot{\vartheta}\rw &=& -{1 \over 3}\vartheta\rw^2 -{4\pi}\kappa\density\rw\ , \label{rayRW} 
\end{eqnarray}
where the ``unperturbed" magnetic field is obviously absent, and the propagation
equations for inhomegeneities
\begin{eqnarray}
\dot{\delta\density}& =& -\vartheta\rw \delta\density-\density\rw \delta\vartheta-\delta\density\delta\vartheta\label{cg2}\\
\dot{\delta\vartheta}&=& -{4\pi}\kappa\delta\density-{2 \over 3}\vartheta\rw\delta\vartheta -\delta\vartheta^2\label{ray2}\\
\dot{H}^i &=&\delta\vartheta (\delta_{\AA}^i \delta^{\AA}_j- \delta_{j}^i )H^j -{2\over 3} \vartheta\rw H^i\label{ind2}\ .
\end{eqnarray}
The perturbations $\delta\density=\density-\density\rw$ and
$\delta\vartheta=\vartheta -\vartheta \rw$ are defined as differences between local and
background values. They are not assumed to be small, hence nonlinear terms in
(\ref{cg2}-\ref{ray2}) remain. The shear tensor for planar
perturbation
\begin{equation}
\sigma_{ik} = {\sigma \over \sqrt{3}} (3\delta_i^\AA\delta_k^\AA - \delta_{ik}) 
= {\delta \vartheta \over {3}} (3\delta_i^\AA\delta_k^\AA - \delta_{ik}) 
\end{equation}
has been employed to derive equations (\ref{ray2} and \ref{ind2}). In particular the
 shear scalar $\sigma$ has been expressed by the variation in the expansion rate
 $\delta\vartheta $, i.e. ${\sigma} = {1\over \sqrt{3}}\delta \vartheta $. The autonomous
 sub-system (\ref{cg2}-\ref{ray2}) can be evaluated to a single second order
differential equation for the density contrast $\x= \rr$
\begin{equation}
\left(\x\right)^{..}-{2\over(1+\x)}\left(\dot{\x}\right)^2
=-{2\over 3}\vartheta\rw\dot{\x}+4\pi\kappa\density\rw\x\left(1+\x\right)
\label{paskudne}
\end{equation}
and solved independently.

\section{The evolution of planar structures}
\label{evolution}

Equation (\ref{paskudne}) 
reduces to the Jeans-Bonnor equation (Bonnor 1957, Weinberg 1972)
\begin{equation}
\left(\ix\right)^{..}
=-{2\over 3}\vartheta\rw\dot{\ix}+4\pi\kappa\density\rw\ix
\label{fajne}
\end{equation}
under the change of the perturbation variable (compare Buchert 1989 in the context of the 
Zeldovich approximation)
\begin{equation}
\ix={\x\left(1+\x\right)^{-1}}\label{nowa} \ .
\end{equation}
This enables one to express the finite amplitude (nonlinear) perturbations as functions
of the solutions to the linear equation (\ref{fajne}). Equation (\ref{fajne}) rewritten
in the conformal time $\eta=\int a(t)^{-1} dt$ takes the form
\begin{equation}
\ix''+{a'\over a}\ix'-{3\over 2}\left[K+\left({a'\over a}\right)^2\right]=0 \ ,
\label{fconf}
\end{equation}
where the dimensionless scale factor $a(\eta)$, formally defined by the differential equation
$\vartheta_0(\eta)'=3 {a(\eta)'\over a(\eta)^2}$, satisfies the Friedman equation for a dust-filled universe
\begin{equation}
{K\over a^2}-{a'\over a^2}+2 {a''\over a^3}=0
\label{Friedman}
\end{equation}
and can be found in the exact form
\begin{equation}
a(\eta) = {{\MM}\over {3 K}} \sin^2\left(\sqrt{K}\eta\over 2\right)\label{radius} \ .
\end{equation}
Prime in (\ref{fconf}), (\ref{Friedman}) and all the equations below is the differentiation with respect to
$\eta$, both $K$ and $\MM=a^3(\eta)\density$ are constants of motion\footnote{In general relativity $K$
means the curvature index and is traditionally set to $K=-1,0,+1$. In Newtonian cosmology $K$ distinguishes
between different dynamical behaviours.}. The two independent solutions of the equation (\ref{fconf}) read 
\begin{eqnarray}
\ix_1(\eta)&=&-10 {\MM\over K}\left[1+{3\over 2}\left(-2+\sqrt{K} \eta \cot{\sqrt{K}\eta\over 2}\right)\sin^{-2}{\sqrt{K}\eta\over 2}\right]\label{mi1}\\
\ix_2(\eta)&=&\left({\MM\over K}\right)^{-3/2}\left[1+\tan^{-2}{\sqrt{K}\eta\over 2}\right]\label{mi2}\cot^3{\sqrt{K}\eta\over 2}
\end{eqnarray}
and in the $K\rightarrow 0$ limit yield $\ix_1(\eta)=\MM \eta^2$ and $\ix_2(\eta)=\MM^{-3/2}\eta^{-3}$,
respectively. The arbitrary solution to (\ref{fajne}) is a~linear combination
\begin{equation}
\ix(\eta)=c_1\ix_1(\eta)+c_2\ix_2(\eta) \ ,
\label{kombinacja}
\end{equation}
while the original perturbation $\x$ can be found by use of the
reciprocal transform
\begin{equation}
\x(\eta)={\ix(\eta)\over{1-\ix(\eta)}} \ .
\label{stara1}
\end{equation}

\noindent 
Although the solutions $\ix(\eta)$ form the linear space (\ref{kombinacja}), the solutions
$\Delta(\eta)$ to equation (\ref{paskudne}) obviously do not. On the other hand the time evolution of
the magnetic field is governed by the linear equation (\ref{ind2}), which after investing (\ref{cg2})
can be integrated exactly. The solution written covariantly reads
\begin{equation}
H^i(\eta)={a^2_{\ini}\over{a^2(1-\ix(\eta))}}(\delta_{j}^i -\ix(\eta)\delta_{\AA}^i \delta^{\AA}_j)H^j_{\ini} 
\label{Hi}
\end{equation}
and can also be expressed as a function of the density contrast $\x(\red)$ of the \pan structure 
observed at the redshift $\red$ (we write it in the matrix form)
\begin{equation}
\left(
\begin{array}{c}
H_x(\red)\\ H_y(\red) \\ H_z(\red)
\end{array}
\right)=
\left({\red+1\over \red_\ini+1}\right)^2
\left(
\begin{array}{ccc}
{1+\x(\red)}& 0 & 0 \\
0 & {1+\x(\red)}& 0 \\
0 & 0 & 1 \\
\end{array}
\right)
\left(
\begin{array}{c}
H_x\\ H_y \\ H_z
\end{array}
\right)_{\ini} \ .
\label{Hv}
\end{equation}
Derivation of the formula (\ref{Hi}-\ref{Hv})  does not involve the Raychaudhuri equation. Therefore,
(\ref{Hi}-\ref{Hv})  is well satisfied for arbitrary large magnetic fields, and is independent of the
$p=0$ approximation. It expresses the amplification of the magnetic field during the one-dimensional
fluid compression in a homogeneously expanding medium. It shows that the orthogonal component $H_z$ is
systematically diluted during the universe expansion, while components parallel to the \pan plane
increase with the local density. As a consequence, the relic magnetic field in a large-scale planar
structure is amplified and flattened to the plane.

All the dynamical effects of gravity, pressure or magnetic field cumulate in the $\ix(\eta)$ or
$\x(\red)$ evolution. For the dust filled universe in the weak magnetic field approximation the formulae
(\ref{Hv}), (\ref{kombinacja}), (\ref{mi1}), (\ref{mi2}) and (\ref{radius}) constitute a~closed form
solution for the magnetic field enhancement.

\section{Numerical estimations}
\label{numerical}

It is reasonable to view the numerical results in the $K=0$ universe,  where the linear solution for
$\ix(\red)$ expressed as a~function of the redshift $\red$ takes the form

\begin{equation}
\ix(\red)={\x_\ini \over 5}\left[3\ksi ^{2} +2\ksi ^{-3}\right]
+{\y_\ini \over 5} \left[\ksi ^{2} -\ksi ^{-3}\right] \ .
\label{kombinacja1}
\end{equation}
Consequently the perturbations $\x=\left[\rr\right]$,  $\y=\left[\ss\right]$  and the magnetic field $H(\red)$ read respectively
\begin{eqnarray}
\x(\red)&=&\frac{2(\x_\ini+3\y_\ini)+3(\x_\ini-2\y_\ini)\ksi^5} {5\ksi^3-2(\x_\ini+3\y_\ini)-3(\x_\ini-2\y_\ini)\ksi^5}\label{xz}\\
\y(\red)&=&\frac{(\x_\ini+3\y_\ini)-(\x_\ini-2\y_\ini)\ksi^5}{5\ksi^3-2(\x_\ini+3\y_\ini)-3(\x_\ini-2\y_\ini)\ksi^5}\label{yz}\\
H(\red)&=&\left({\red+1\over \red_\ini+1}\right)^2 (1+\x(\red)) H_\ini\label{Hz} \ .
\end{eqnarray}
Formulae (\ref{xz}-\ref{Hz}) show how the structure formation process
depends on the initial conditions at the recombination, i.e. on the initial density fluctuation $\x_\ini$ and the initial flow $\y_\ini$ 
(the velocity divergence). Setting $\x_\ini=10^{-6}$ at $\red=1200$, as suggested by the CMBR measurements, we draw
$\x$ and $\y$ as functions of $\red$ and the initial flow $\y_\ini$ (see Fig. 1 and Fig. 2).

\begin{figure}[h]
\centering
\psfig{file=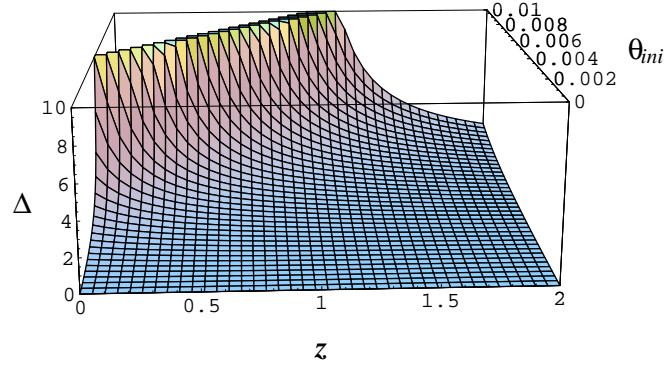, width=88mm }
\caption{Density contrast as a function of redshift $z$ and initial velocity flow $\y_\ini$.}
\label{Fig1}
\end{figure}

\begin{figure}[h]
\centering
\psfig{file=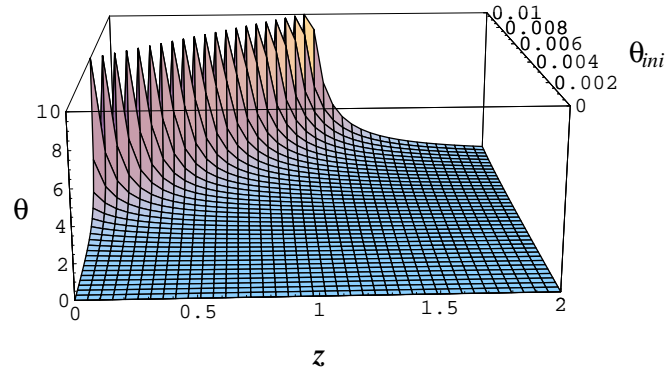, width=88mm }
\caption{Velocity field contrast as a function of redshift $z$ and initial velocity flow $\y_\ini$.}
\label{Fig2}
\end{figure}
Galactic superclusters with the density contrast $\x \leq 100$  are considered as nonvirialized
structures (Peacock 1999). Their length scales of about $\sim 10$~Mpc are typically expected at the moment
of transition from the linear to nonlinear regime. The evolution of these structures relative to the
initial values of the large scale inflows $\y_\ini$ is presented in Fig
3. The evolutionary paths with different  $\y_\ini$  form horizontal lines, while the solid sloped
lines mark the beginning ($\x = 1$, $\y=-0.35$) and the end ($\x = 10$, $\y=-3$) of the 
{\wnr} (the shaded region).  The magnitudes of the contrasts at both characteristic
moments are set to be compatible with the values obtained  from the numerical simulation (Gramann et al. 2002). 
The region below the shaded region represents  the linear evolution, while the region
above --- the strongly nonlinear collapse. The existence of the low $\red$ structures with $\x \leq
100$, confirmed by the observations, when compared with the theoretical low redshift   behaviour
(Fig 3) favours the initial inflow $\y_\ini\simeq 10^{-3}$ at the recombination
epoch\footnote{Note that the fluid velocities and not their divergences contribute to the Sachs-Wolfe
temperature formula, therefore $\y_\ini$ is not directly determined from the CMBR satellite
measurements}.

\begin{figure}[h]
\centering
\psfig{file=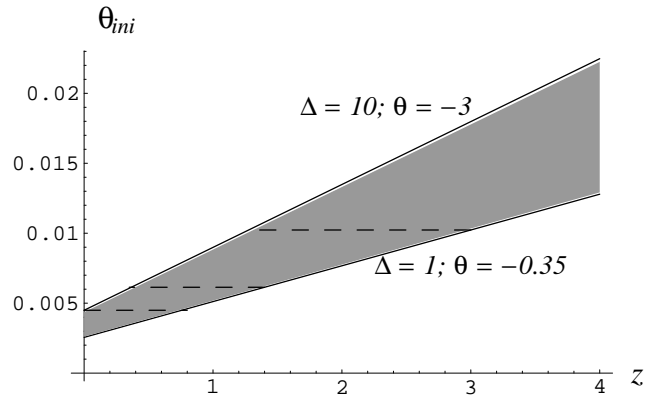, width=88mm }
\caption{ Schematic diagram of weakly nonlinear evolution of the density and velocity contrasts.
 For both contrasts the $\x = const$ and $\y = const$
  sections  are presented. The solid lines correspond to the initial and final 
 states of the previrial nonlinear evolution of  $\x $ and $\y$. 
The shaded area represents all the evolutionary states between ($\x = 1$, $\y=-0.35$) and ($\x = 10$, $\y=-3$). 
 Horizontal dashed lines show the evolutionary tracks.}
\label{Fig3}
\end{figure}
The initial velocity field determines the  basic time scales for the weakly and strongly nonlinear
structure formation respectively, and consequently, defines the time scales for compression and
magnification of the seed magnetic field. In particular, the time of entering the strongly nonlinear regime 
($\x\approx 150$)  is crucial for the efficiency of the dynamo action (e.g. Widrow 2002).  The amplification degree in this
process exponentially depends on the time available for the dynamo mechanism to operate. The pre-dynamo
magnetic field  compression (Lesch et al. 1995) is followed  by the dynamo amplification, which in the case of
protogalaxies results in some $10^4- 10^{10}$ enhancement in amplitude, depending on the time when the
field reaches the observed value. The question if the dynamo mechanism is applicable to galactic
cluster structures is still open. Even more intriguing is the origin of the almost equipartized
supercluster magnetic fields ($\simeq 10^{-7}$) (Sigl et al. 1998). Fig. 3 illustrates that structures with
initial inflow $\y_\ini \simeq 10^{-2}$ reaches the {\wnr} ($\x \simeq 1$) at redshift $\red=3$, and
leaves the weakly nonlinear phase ($\x \simeq 100$) at $\red=1.3$. A dynamo may operate then for $\simeq
8\times10^9 yr$, which (taking into account the typical dynamo time scale $\simeq 10^9 yr$) results in the
$\simeq 2\,10^3$ enhancement of magnetic field. On the other hand, for the structures with the initial
inflows at the level of $\y_\ini \approx 6\times 10^{-3}$ (or $\y_\ini \approx 5\times 10^{-3}$), the beginning
and the end of the previrial nonlinear stage occur at $\red \approx 1.4$ (\red ~$\approx 0.8$) and $\red
\approx 0.4$ ($\red \approx 0$), respectively, which limits the dynamo operation phase to less than
$2.3\times 10^9yr$ (or none). In this latter case i.e. for systems which presently attain the previrial
collapse state ($\x \simeq 100$ at $z\simeq 0$), the primordial magnetic field constrained at $z = 10^3$
to magnitudes below $\sim 10^{-9}$ G (Barrow et al. 1997) are  pre-dynamo amplified by merely 2 orders of magnitude.

\section{Summary}

We provided  the nonlinear exact solutions for the density,  velocity and magnetic fields 
for the \pan-type structures in the Newtonian expanding universe. The approximations of the potential velocity field 
and vanishing matter pressure have been employed. The time when the compressing flat structure enters the regime of
nonlinear growth is controlled by the initial value of velocity field at the recombination. The structures accompanied 
by large hydrodynamic flows collapse earlier, i.e. the moment when dynamo mechanism
may switch on occurs at higher redshifts, which eventually results in stronger magnetic field enhancement.
For presently observed velocity fields, $10^{-1}$, (see, e.g. Dekel 1997) in supergalactic structures of $100$~Mpc and $\x \sim 1$ 
the initial inflows $\y_\ini \simeq 10^{-3}$ and initial magnetic fields $H_\ini\simeq 10^{-9} - 10^{-8}$~Gauss are expected. 
The result is compatible with the simulation estimations (Gramann et al., 2002). 

Firm evidence of primordial magnetic fields in structures at pre-virial stages is of particular importance, as these
fields "remember" the initial conditions and thus set constraints on the seed magnetic fields, density and
velocity fields at the recombination. The observational techniques become more important 
(Faraday rotation measurements and the indications coming from the propagation of cosmic radiation 
UHE in the Local Supercluster), which potentially might distinguish between the large scale magnetic 
seed component from other magnetic fields of astrophysical origin (i.e. resulting from galactic dynamo, outflows from 
radiogalaxies etc.). The rotation measure which have the same sign along the Supercluster plane would suggest 
a coherent, relic field at this scale.

\section{Acknowledgments}
We thank the Referee for efforts to improve the manuscript.

\section*{References}

\parindent=7mm
\noindent
Bagchi, J.,Ensslin, T., Miniati, F., et al. 2002, New Astron., 7, 249\\
Barrow, J., Ferreira, P., $\&$ Silk, J. 1997, Phys. Rev. Lett., 78, 3610\\
Bonnor, W. 1957, MNRAS, 117, 104\\
Buchert, T. 1989, A$\&$A, 223, 9\\
Chandrasekhar, S., $\&$ Fermi, E. 1953, ApJ, 118, 116\\
Dekel, A. 1997, in {\em Formation of Structure in the Universe}, ed. A. Dekel,  J. Ostriker (Cambridge Univ. Press)\\
Ellis, G. 1971, in {\it Lectures in General Relativity and Cosmology},  
Proc. of International School of Physics Enrico Fermi,  XLVII, ed. R. Sachs\\
Gramann, M., $\&$ Suhhonenko, I. 2002, MNRAS, 337, 1417\\
Kim, E., Olinto, A., $\&$ Rosner, R. 1996 ApJ, 468, 28\\
Kim, K., Tribble, P., $\&$ Kronberg, P. 1991, ApJ, 379, 80\\
Kronberg, P. 1994, Rep. Prog. Phys., 57, 325\\
Lesch, H., $\&$ Chiba, M. 1995, A$\&$A, 297, 305\\
Papadopoulos, D., $\&$ Esposito, F. 1982, ApJ, 257, 10\\
Peacock, J. 1999, {\em Cosmological Physics}, (Cambridge Univ.Press)\\
Peebles, P. 1993, {\em Principles of Physical Cosmology}, (Princeton Univ.Press)\\
Sigl, G., Lemoine, M., $\&$ Biermann, P. 1999, Astropart. Phys., 10, 141\\
Tsagas, C., $\&$ Barrow, J. 1997, Class. Quantum Grav., 14, 2539\\
Wasserman, I. 1978, ApJ, 224, 337\\
Weinberg, S. 1972, {\it Gravitation and Cosmology}, (John Wiley and Sons Inc. New York)\\
Widrow, L. 2002, Rev. of Modern Phys., 74, 775\\

\end{document}